\documentclass[aps,prl,showpacs,floatfix,twocolumn,amsmath,amssymb,preprintnumbers]{revtex4-1}
\usepackage{mathrsfs}
\usepackage[figuresright]{rotating}
\usepackage{amsmath}
\usepackage{amssymb}
\usepackage{graphicx}% Include figure files
\usepackage{color}
\usepackage{dcolumn}% Align table columns on decimal point
\usepackage{bm}% bold math
\usepackage[breaklinks=true,colorlinks=true,linkcolor=blue,urlcolor=blue,citecolor=blue]{hyperref}

\usepackage{soul}

\makeatletter
\newcommand{\rmnum}[1]{\romannumeral #1}
\newcommand{\Rmnum}[1]{\expandafter\@slowromancap\romannumeral #1@}
\makeatother

\begin{document}

\title{Synthesis and physical properties of CeRu$_2$As$_2$ and CeIr$_2$As$_2$}

\author{Kangqiao Cheng$^{1}$}
\author{Xiaobo He$^{1}$}
\author{Haiyang Yang$^{2}$}
\author{Binjie Zhou$^{1}$}
\author{Yuke Li$^{2}$}
\email[]{yklee@hznu.edu.cn}
\author{Yongkang Luo$^{1}$}
\email[]{mpzslyk@gmail.com}
\address{$^1$Wuhan National High Magnetic Field Center and School of Physics, Huazhong University of Science and Technology, Wuhan 430074, China;}
\address{$^2$Department of Physics and Hangzhou Key Laboratory of Quantum Matters, Hangzhou Normal University, Hangzhou 311121, China.}

\date{\today}

\begin{abstract}

We studied the physical properties of two Kondo-lattice compounds, CeRu$_2$As$_2$ and CeIr$_2$As$_2$, by a combination of electric transport, magnetic and thermodynamic measurements. They are of ThCr$_2$Si$_2$-type and CaBe$_2$Ge$_2$-type crystalline structures, respectively. CeRu$_2$As$_2$ shows localized long-range antiferromagnetic ordering below $T_N$=4.3 K, with a moderate electronic Sommerfeld coefficient $\gamma_0$=35 mJ/mol$\cdot$K$^2$. A field-induced metamagnetic transition is observed near 2 T below $T_N$. Magnetic susceptibility measurements on aligned CeRu$_2$As$_2$ powders suggest that it has an easy axis and that the cerium moments align uniaxially along $\mathbf{c}$ axis. In contrast, CeIr$_2$As$_2$ is a magnetically nonordered heavy-fermion metal with enhanced $\gamma_0$$>$300 mJ/mol$\cdot$K$^2$. The initial onset Kondo temperatures of the two compounds are respectively 6 K and 30 K. We discuss the role of the crystal structure to the strength of Kondo coupling. This work provides two new dense Kondo-lattice materials for further investigations on electronic correlation, quantum criticality and heavy-electron effects.

\end{abstract}

%\pacs{75.75.-c, 72.55.+s, 75.30.Kz, 75.10.-b}
%75.75.-c   Magnetic properties of nanostructures
%72.55.+s 	Magnetoacoustic effects
%75.30.Kz 	Magnetic phase boundaries (including magnetic transitions, metamagnetism, etc.)
%75.10.-b 	General theory and models of magnetic ordering

\maketitle

\section{\Rmnum{1}. Introduction}

The hybridization ($J_{cf}$) between conduction ($c$) electrons and more localized $f$ electrons in Kondo-lattice compounds simultaneously yields two competing phenomena: the Ruderman-Kittel-Kasuya-Yosida (RKKY) interaction\cite{RKKY-RK,RKKY-K,RKKY-Y} and the Kondo effect\cite{Kondo-RMinimum,Hewson-Kondo}. While the RKKY interaction mediates the magnetic exchange between local moments and stabilizes a long-range magnetic ordering, the consequence of Kondo effect is to screen and quench the local moments. Depending on the strength of $J_{cf}$, the ground state of Kondo lattices varies from localized magnetic ordered regime for small $J_{cf}$, heavy-fermion regime for moderate $J_{cf}$ to intermediate-valence regime for large $J_{cf}$.

\begin{figure}[!ht]
\vspace*{-15pt}
\hspace*{-15pt}
\includegraphics[width=10.0cm]{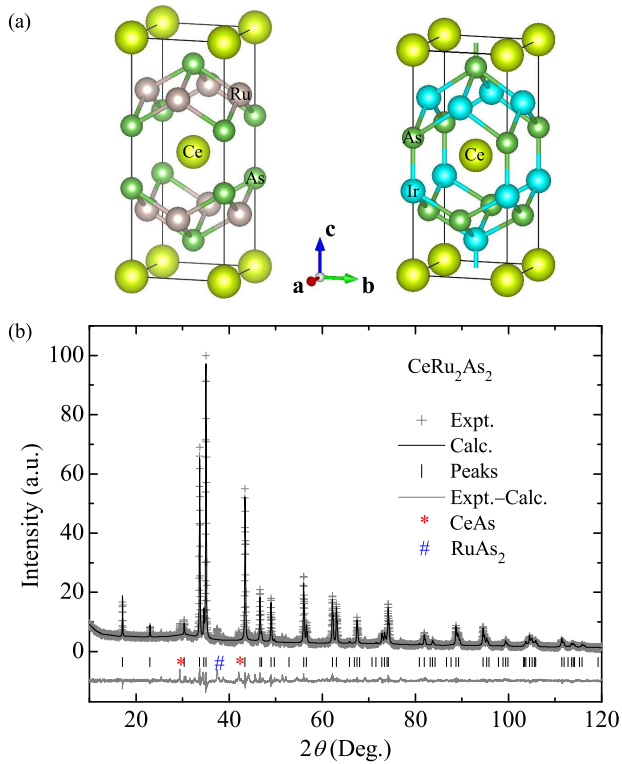}
\vspace*{-25pt}
\caption{\label{Fig1} (a) Crystalline structure of ThCr$_2$Si$_2$-type CeRu$_2$As$_2$ (left) and CaBe$_2$Ge$_2$-type CeIr$_2$As$_2$ (right). (b) Rietveld refinement of CeRu$_2$As$_2$ XRD pattern.}
\end{figure}

``Ce-122" refers to a big family of Kondo-lattice compounds. It is mainly comprised of two types of tetragonal crystalline structures: the ThCr$_2$Si$_2$-type (I4/$mmm$, No. 139) and CaBe$_2$Ge$_2$-type (P4/$nmm$, No. 129). We show their crystalline structures in Fig.~\ref{Fig1}(a). The discovery of superconductivity in K-doped BaFe$_2$As$_2$ rekindled the interest in the ThCr$_2$Si$_2$ structure\cite{Rotter-BaK122SC}. In this structure, two vertically inverted CrSi layers are alternatingly sandwiched with Th ions embedded in between. Historically, ThCr$_2$Si$_2$ structure was also well-known for hosting a number of Kondo-lattice materials (see reviews \cite{Stewart-RMPNFL,Misra-HFSystem,Pfleiderer-RMPHFSC}), including CeCu$_2$Si$_2$ - the first heavy-fermion superconductor\cite{Steglich-CeCu2Si2SC}, URu$_2$Si$_2$ - hidden order superconductor\cite{Palstra-URu2Si2HO}, YbRh$_2$Si$_2$ - Kondo breakdown quantum critical point\cite{Custers-YbRh2Si2QCP} \textit{et al}. The CaBe$_2$Ge$_2$-type structure is relatively less famous. In this structure, the Be and Ge sites are interchanged in every other layer, and interlayer couplings can be bridged by Be-Ge bonding, which renders a more three-dimensional network than ThCr$_2$Si$_2$. Some examples are CeRh$_2$P$_2$\cite{Madar-CeRh2P2}, CeNi$_2$As$_2$\cite{Ghadraoui-LnNi2As2}, CeIr$_2$Si$_2$\cite{Hiebl-CeT2Si2} \textit{etc}. Note that CeNi$_2$As$_2$ can crystallize in both structures\cite{Ghadraoui-LnNi2As2,Suzuki-CeNi2X2,LuoY-CeNi2As2}. For whatever ThCr$_2$Si$_2$- or CaBe$_2$Ge$_2$-type Ce-122, the Ce-$4f$ electrons interact with the conduction electrons donated by CrSi or BeGe layers, and such a $c$-$f$ hybridization builds up a natural platform to investigate Kondo effect and electronic correlations. A variety of interesting emergent states have been observed in Ce-122 compounds, \textit{e.g.} metamagnetism, heavy-fermion, non-Fermi liquid, quantum critical point (QCP), and unconventional superconductivity\cite{Flouquet-CeRu2Si2MM,LuoY-CeNi2As2,Steglich-CeCu2Si2SC,Yuan-CeCu2Si2TwoSC,Grosche-CePd2Si2SC,Movshovich-CeRh2Si2SC,LuoY-CeNi2As2Pre}, the mechanisms of which remain controversial. In particular, local centrosymmetry is broken in CaBe$_2$Ge$_2$-type Ce-122, making them candidates to look for new non-centrosymmetric heavy-fermion superconductors. Extensive material bases are required in this field.

In this paper, we study two Ce-122 Kondo-lattice compounds, CeRu$_2$As$_2$ and CeIr$_2$As$_2$. The ThCr$_2$Si$_2$-type CeRu$_2$As$_2$ is a new material. The synthesis of CaBe$_2$Ge$_2$-type CeIr$_2$As$_2$ has been reported by Pfannenschmidt \textit{et al}\cite{Pfannenschmidt-ReIr2Pn2}, but its physical properties have not been well studied. Our work reveals that CeRu$_2$As$_2$ is a $4f$-electron localized antiferromagnet with N\'{e}el temperature $T_N$=4.3 K and a moderate Sommerfeld coefficient $\gamma$=35 mJ/mol$\cdot$K$^2$, whereas CeIr$_2$As$_2$ is magnetically nonordered heavy-fermion metal with enhanced Sommerfeld coefficient $\gamma_0$$>$300 mJ/mol$\cdot$K$^2$. The initial onset Kondo temperatures of the two compounds are respectively 6 K and 30 K. These results place them in the regimes of magnetic ordered with small $J_{cf}$ and heavy-fermion with moderate $J_{cf}$, respectively.

\section{\Rmnum{2}. Experimental details}

Polycrystalline CeRu$_2$As$_2$ and CeIr$_2$As$_2$ samples were synthesized by the method of solid-state reaction. High-purity Ce, Ru, Ir and As were used as starting materials. First, CeAs, RuAs and IrAs were prepared by reacting As powders with Ce, Ru and Ir powders at 973 K, 1073 K and 1273 K respectively for 24 hours. Then, powders of CeAs, RuAs and Ru were weighted according to the stoichiometric ratio, thoroughly ground and pressed into
a pellet under a pressure of 600 MPa in an argon-filled glove box. The pellet was packed into an alumina crucible and sealed into an evacuated quartz tube, which was then slowly heated to 1323 K and kept at that temperature for 48 hour. After that, the resultant was re-ground and re-sintered for two more times to achieve a good homogeneity. The synthesis for CeIr$_2$As$_2$ was essentially similar, while the sintering temperature was at 1373 K. The non-$4f$ analogs LaRu$_2$As$_2$ and LaIr$_2$As$_2$ were also grown, in the same method.

Powder x-ray diffraction (XRD) patterns were recorded at room temperature on a PANalytical X-ray diffractometer with Cu K$\alpha$ radiation. Electrical resistivity was measured by standard four-probe method in a physical property measurement system (PPMS-9, Quantum Design), which was also used for the specific heat measurements. Magnetization measurements were performed using a magnetic property measurement system (MPMS-VSM, Quantum Design). The measurements were made after a zero-field-cooling (ZFC) process.

\section{\Rmnum{3}. Results and Discussion}

The crystalline structures of CeRu$_2$As$_2$ and CeIr$_2$As$_2$ are shown in Fig.~\ref{Fig1}. For the new compound CeRu$_2$As$_2$, we performed the Rietveld refinement to the XRD. The peaks are well indexed to the tetragonal ThCr$_2$Si$_2$-type structure, except for a little impurity phase of CeAs and RuAs$_2$. The best fitting parameters are $a$=4.1696(5) \AA, $c$=10.3868(7) \AA, and the atomic coordinate of As (0, 0, 0.3692(4)). More details about the structural parameters of CeRu$_2$As$_2$ and CeIr$_2$As$_2$ can be found in Table ~\ref{Tab1}. The lengths of $a$- and $c$-axes of CeIr$_2$As$_2$ obtained in this work are close to but a little smaller than those in literature\cite{Pfannenschmidt-ReIr2Pn2}. The Ir-As bonding ($d_{\text{Ir-As}}$=2.33 \AA) between adjacent IrAs layers makes $c$-axis much shorter than in CeRu$_2$As$_2$ ($d_{\text{As-As}}$=2.72 \AA).

\begin{table}[ht]
\tabcolsep 0pt \caption{\label{Tab1} Crystallographic parameters of CeRu$_2$As$_2$ from the Rietveld refinement to the Powder X-ray Diffractions at 300 K. The data of CeIr$_2$As$_2$ are also shown for comparison. }
\vspace*{-15pt}
\begin{center}
\def\temptablewidth{1\columnwidth}
{\rule{\temptablewidth}{1pt}}
\begin{tabular*}{\temptablewidth}{@{\extracolsep{\fill}}ccc}
    Compounds            &  CeRu$_2$As$_2$  &  CeIr$_2$As$_2$  \\   \hline
    Space group          &  $I4/mmm$        &    $P4/nmm$      \\
    $a$ (\AA)            &  4.1696(5)          &  4.2865(6)          \\
    $c$ (\AA)            &  10.3868(7)         &  9.8849(9)          \\
    $V$ (\AA$^{3}$)      &  180.580         &  181.625         \\
    $Z$                  &  2               &  2               \\
    $\rho$ (g/cm$^{3}$)  &  9.050           &  12.332          \\
    $R_{\text{wp}}$ (\%) &  9.36            &  11.57           \\
\end{tabular*}
{\rule{\temptablewidth}{1pt}}
\end{center}
\end{table}

\begin{figure*}[t]
\vspace*{-20pt}
\hspace*{0pt}
\includegraphics[width=16cm]{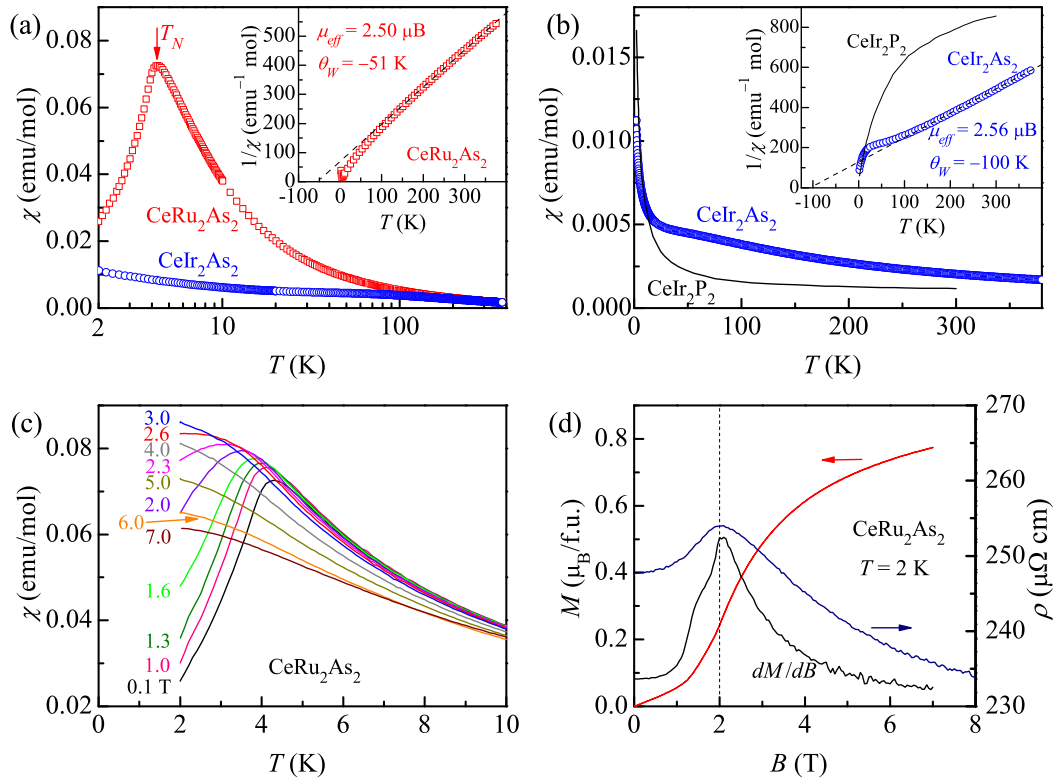}
\vspace*{-20pt}
\caption{\label{Fig2} (a-b) Temperature dependent magnetic susceptibility of polycrystalline CeRu$_2$As$_2$ and CeIr$_2$As$_2$. CeRu$_2$As$_2$ undergoes an AFM transition at $T_N$=4.3 K. The insets display the Curie-Weiss fittings. The curve of CeIr$_2$P$_2$ (black line) reproduced form Ref.~\cite{Pfannenschmidt-ReIr2Pn2} is shown for comparison. (c) Evolution of the AFM peak in $\chi(T)$ of CeRu$_2$As$_2$ under various fields. (d) Field dependent magnetization (left), resistivity (right) and derivative susceptibility ($dM/dB$) of CeRu$_2$As$_2$ at 2 K.}
\end{figure*}

Figure \ref{Fig2}(a-b) show the temperature dependence of magnetic susceptibility ($\chi$=$M/B$) of polycrystalline CeRu$_2$As$_2$ and CeIr$_2$As$_2$. For temperature above 150 K, $\chi$ of both compounds obey standard Curie-Weiss formula, $\chi(T)$=$\frac{C}{T-\theta_W}$, where $\theta_W$ is the Weiss temperature. The fittings yield the effective moment $\mu_{eff}$=2.50 and 2.56 $\mu_B$ for CeRu$_2$As$_2$ and CeIr$_2$As$_2$, respectively, very close to that of a free Ce$^{3+}$ ion, 2.54 $\mu_B$. This implies that the Ru and Ir ions are essentially nonmagnetic. The derived $\theta_W$ is $-$51 K for CeRu$_2$As$_2$, suggestive of antiferromagnetic correlation among Ce moments. A hump is seen near 80 K in CeIr$_2$As$_2$, which makes $\chi$ less temperature-dependent at low temperature. Similar behavior was also seen in CeCoIn$_5$ and CeIrIn$_5$\cite{Petrovic-CeCoIn5SC}, which is probably due to the crystalline electric field (CEF) effect, see below. For comparison, the magnetic susceptibility of CeIr$_2$P$_2$ (data reproduced from Ref.~\cite{Pfannenschmidt-ReIr2Pn2}) is also shown in Fig.~\ref{Fig2}. Pfannenschmidt \textit{et al} placed CeIr$_2$P$_2$ in the regime of intermediate-valence, based on the Pauli-paramagnetic-like $\chi(T)$. Apparently, the ``chemical pressure" effect of As-P substitution greatly enhances the $c$-$f$ hybridization, and the $4f$ electrons become delocalized in CeIr$_2$P$_2$. At low temperature, the most prominent feature of CeRu$_2$As$_2$ is that $\chi(T)$ displays a sharp peak at 4.3 K, manifesting an antiferromagnetic (AFM) transition which will be studied further. No anomaly is seen in CeIr$_2$As$_2$ at low $T$; the slight increase in $\chi$ is probably due to some magnetic impurities (CeAs). We should point out that the fitted $\theta_W$ from the high$-T$ region is $\sim-$110 K for CeIr$_2$As$_2$, much larger than that of CeRu$_2$As$_2$. Since CeIr$_2$As$_2$ does not show any magnetic ordering, this enhanced $\theta_W$ is likely promoted by the Kondo effect. Further transport and specific heat measurements suggest that CeIr$_2$As$_2$ sits much closer to a QCP.

\begin{figure}[t]
\vspace{-5pt}
\hspace{-10pt}
\includegraphics[width=9.0cm]{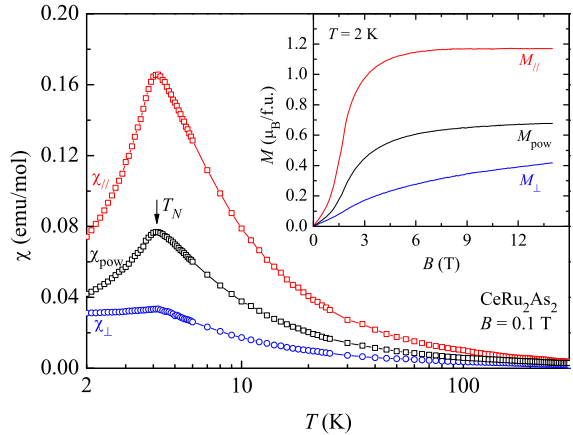}
\vspace{-25pt}
\caption{\label{Fig3} Temperature dependence of $\chi_{\parallel}$ and $\chi_{\perp}$ measured from aligned CeRu$_2$As$_2$ powders. The inset shows the isothermal field dependent magnetization. $\parallel$ and $\perp$ are with respect to the aligning field $\mathbf{B_{al}}$. Seeing Appendix. The results for unaligned powders ($\chi_{\text{pow}}$) are shown for comparison.}
\end{figure}

In Fig.~\ref{Fig2}(c), we display $\chi(T)$ of CeRu$_2$As$_2$ under different magnetic fields. As field increases, the peak in $\chi(T)$ is gradually suppressed, and meanwhile the peak position also moves to lower temperature, characteristic of an AFM transition. For field at about 2.6 T, the peak disappears, and $\chi(T)$ saturates at low temperature. Further increasing $B$, the value of magnetic susceptibility decreases systematically, and no clear anomaly is seen in $\chi(T)$ only except for a trend of saturation at low $T$. Such an evolution with field typically entails a field-induced metamagnetic transition, as observed in many cerium compounds\cite{Mun-CeGeNi3,LuoY-CeNi2As2}. Since Ce$^{3+}$ has a small de Gennes factor\cite{Blundell-MaginCM}, the magnetic exchange coupling between cerium moments are generally weak, thus, the magnetic moments can be reoriented by a moderate field. This is indeed the case in CeRu$_2$As$_2$. In Fig.~\ref{Fig2}(d), we present isothermal field dependent magnetization and derivative susceptibility ($dM/dB$) at 2 K. Under low field, the magnetization increases linearly. A speed-up is visible for field larger than 1 T, and finally tends to saturate above 4 T. This trend is more clearly seen in $dM/dB$ that peaks near 2 T. Similar metamagnetic transition was interpreted as a spin-flop in CeNi$_{2}$As$_2$ where the magnetic moments are uniaxially aligned along $\mathbf{c}$ and a tiny hysteresis is seen near the metamagnetic transition\cite{LuoY-CeNi2As2}. In CeRu$_2$As$_2$, the hysteresis is negligible, probably due to the polycrystalline sample.

To further study the magnetic anisotropy of CeRu$_2$As$_2$, we measured the susceptibility of field-aligned powders. The polycrystalline sample is thoroughly ground into powders (This process is carried out carefully in a glove box), and mixed with Stycast 1266 epoxy with a small weight ratio $\sim$0.2 so that the grains are well isolated by epoxy\cite{Young-Alignment}. The mixture is then placed in a strong aligning field ($\mathbf{B_{al}}$) of 14 T at 300 K in PPMS and held motionless for 12 hours before the Stycast is completely cured. The magnetic susceptibility for parallel ($\chi_{\parallel}$) and perpendicular ($\chi_{\perp}$) to $\mathbf{B_{al}}$ are shown in Fig.~\ref{Fig3} as a function of $T$. (Refer to Appendix for more details.) The important findings are: (\rmnum{1}) $\chi_{\perp}$ is much smaller than $\chi_{\parallel}$, indicative of strong magnetic anisotropy. In particular, we notice the value of $\chi_{\perp}$ is less than a half of $\chi_{\parallel}$ over the full range 2-300 K, for instance, at 300 K, $\chi_{\parallel}$=3.65$\times$10$^{-3}$ emu/mol, and $\chi_{\perp}$=1.66 $\times$10$^{-3}$ emu/mol. This suggests that the compound has an easy axis, rather than easy plane, because for the latter case, $\chi_{\parallel}$=$\chi_{a,b}$, and $\chi_{\perp}$=$(\chi_{a,b}+\chi_c)/2$ (Seeing Appendix); $\chi_{\perp}$$<$$\chi_{\parallel}/2$ is not likely for cerium-contained compounds. (\rmnum{2}) At low temperature, a sharp peak is seen at $T_N$ in $\chi_{\parallel}$, characteristic of an AFM transition. In contrast, the peak in $\chi_{\perp}$ is very shallow, and the susceptibility tends to saturate below $T_N$. These features are highly suggestive that in the AFM state the Ce moments are uniaxially along the easy axis. (\rmnum{3}) The inset to Fig.~\ref{Fig3} shows field dependent magnetization of aligned powders at 2 K. $M_{\parallel}$ increases with field rapidly and saturates at 1.17 $\mu_B$ under high magnetic field. The metamagnetic transition is obviously seen in $M_{\parallel}$, even sharper than in unaligned powders ($M_{\text{pow}}$). $M_{\perp}$ is much smaller than $M_{\parallel}$ and keeps increasing at 14 T. A little sign of metamagnetic transition is also visible in $M_{\perp}$, which probably arises from a small portion of unaligned powders. In light of these observations, it is reasonable to propose $\mathbf{c}$ as the easy axis of CeRu$_2$As$_2$, and below $T_N$ the magnetic moments also align antiferromagnetically along $\mathbf{c}$. This makes a field-induced spin-flop possible. In fact, the $dM/dB$ in Fig.~\ref{Fig2}(d) shows a small shoulder before 2 T, which potentially points to a secondary transition prior to the polarized paramagnetic state. Given an easy-axis antiferromagnet, this secondary transition might be a signature for an intermediate spin-flop phase. Of course, to get a precise magnetic structure, high-quality single crystals and microscopic measurements like neutron scattering experiments are needed. Crystalline electric field (CEF) effect often plays a key role to such a magnetic anisotropy in cerium compounds.

\begin{figure}[t]
\vspace{-5pt}
\hspace{-10pt}
\includegraphics[width=9.0cm]{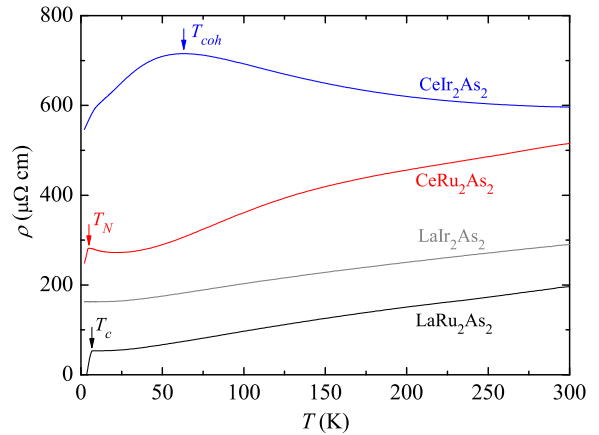}
\vspace{-25pt}
\caption{\label{Fig4} Temperature dependent electric resistivity of CeRu$_2$As$_2$, CeIr$_2$As$_2$, LaRu$_2$As$_2$ and LaIr$_2$As$_2$.}
\end{figure}

The resistivity ($\rho$) of CeRu$_2$As$_2$, CeIr$_2$As$_2$ and their La-counterparts are presented in Fig.~\ref{Fig4}. A previous work on LaRu$_2$As$_2$ by Guo \textit{et al} has revealed metallic behavior and a superconducting transition at $T_c$=7.8 K\cite{Guo-LaRu2As2SC}, and these features are well reproduced in the current work. For CeRu$_2$As$_2$, $\rho(T)$ shows a hump around 150 K, which should be a consequence of CEF splitting of Ce$^{3+}$ $j$=5/2 multiplet. $\rho(T)$ slightly turns up below 22 K, and then decreases sharply below 4.3 K, reminiscent of reduction in spin scattering due to the formation of long-range AFM ordering. LaIr$_2$As$_2$ behaves like a simple metal, say, upon cooling down, $\rho(T)$ decreases almost linearly above 100 K, and then tends to flatten by showing some $T^2$-like behavior at low temperature, characteristic of Fermi liquid. The large residual resistivity $\rho_0$ is probably because of the sample quality. In contrast, CeIr$_2$As$_2$ exhibits the typical dense Kondo behavior (see \textit{e.g.} CeIrIn$_5$\cite{Petrovic-CeIrIn5SC,Takaesu-CeIrIn5Pre}): $\rho(T)$ initially increases as $T$ decreases, and then turns down rapidly after passing through a broad peak near $T_{coh}$=63 K. $T_{coh}$ designates a crossover from incoherent Kondo scattering regime for $T$$>$$T_{coh}$ where Ce moments behaves like separate single-ion impurities to the coherent Kondo scattering regime for $T$$<$$T_{coh}$ where the Ce-$4f$ electrons develop strongly-correlated bands. At low temperature, $\rho(T)$ keeps decreasing linearly down to 2 K, the base temperature of our measurements, without restoring any signature of Fermi-liquid behavior. This places CeIr$_2$As$_2$ in a regime of non-Fermi liquid or ``strange metal"\cite{Stewart-RMPNFL} which is usually observed in the vicinity of a quantum critical point\cite{Lohneysen-CeCu6Au2001,Custers-YbRh2Si2QCP,Park-CeRhIn5Iso,Custers-Ce3Pd20Si6QCP,LuoY-CeNiAsOQCP}. Sub-Kelvin measurements are needed in the future to further clarify this issue.

For CeRu$_2$As$_2$, we also took a field dependent resistivity measurement at 2 K (below $T_N$), as shown in Fig.~\ref{Fig2}(d). Under low field, $\rho$ increases with field, because external magnetic field disturbs the long-range AFM ordering and causes more spin scattering. A maximum is seen in $\rho(B)$ at 2 T, and after that, the resistivity decreases again. The critical field 2 T is coincident with the field at which $M(B)$ shows the largest slope and $dM/dB$ peaks. Further increasing $B$, spin scattering is reduced as the moments are gradually polarized, and thus the resistivity decreases.

\begin{figure}[t]
\vspace{-15pt}
\hspace{-10pt}
\includegraphics[width=9.0cm]{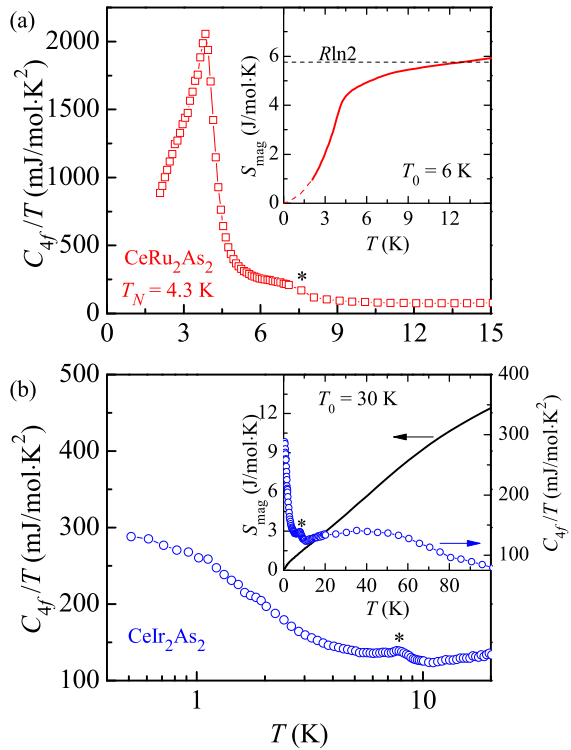}
\vspace{-25pt}
\caption{\label{Fig5} The $4f$-electron contribution to specific heat divided by temperature, $C_{4f}/T$, as a function of $T$ of CeRu$_2$As$_2$ (a) and CeIr$_2$As$_2$ (b). The insets show the magnetic entropy derived by integrating $C_{4f}/T$ over $T$. A broad maximum seen in $C_{4f}/T$ of CeIr$_2$As$_2$ near 40 K indicates $\sim$90 K of the first CEF splitting, which is consistent with what is seen in $\chi(T)$. The ``*" in both panels designate the anomalies due to CeAs impurity.}
\end{figure}

Turning now to the specific heat. For both CeRu$_2$As$_2$ and CeIr$_2$As$_2$, we respectively subtract the specific heat of LaRu$_2$As$_2$ (measured under field of 1 T) and LaIr$_2$As$_2$, and the resultants are the contribution from $4f$ electrons, $C_{4f}$. Fig.~\ref{Fig5}(a) shows $C_{4f}/T$ of CeRu$_2$As$_2$ as a function of $T$. A $\lambda$-shape peak is clearly seen at the transition temperature $T_N$, manifesting a second-order phase transition. A small sub-structure is seen between 6-9 K, which should be from some CeAs impurity that undergoes an AFM transition at 7.6 K\cite{Suzuki-CeAsAFM}. The Sommerfeld coefficient $\gamma_0$ estimated from the paramagnetic state is 35 mJ/mol$\cdot$K$^{2}$, in line with the well localized and ordered $4f$ electrons. We calculated the magnetic entropy ($S_{\text{mag}}$) by integrating $C_{4f}/T$ over $T$. For the low temperature part, we have linearly extrapolated the $C_{4f}/T$ to $T$$\rightarrow$0 limit to ensure $S_{\text{mag}}(0)$=0, and the result is plotted in the inset to Fig.~\ref{Fig5}(a). The entropy gain is about 73\% $R\ln2$ at $T_N$, and fully recovers $R\ln2$ at 12 K. The initial onset Kondo temperature can be estimated as $T_0$$\simeq$6 K through a widely accepted criterion $S_{\text{mag}}(T_0/2)$=0.4$R\ln2$ \cite{Gegenwart2008}. For CeIr$_2$As$_2$, $C_{4f}/T$ turns up logarithmically at low temperature, and tends to level off below 1 K. This suggests that the Fermi liquid behavior likely restores at low temperature with greatly enhanced quasiparticle effective mass and Sommerfeld coefficient $\gamma_0$$>$300 mJ/mol$\cdot$K$^{2}$. Such kind of behavior is usually seen in systems beyond but close to a QCP\cite{Bruning-CeFePOFMHF,LuoY-CeFeAsPO,WangL-CeCo2Ga8}. The estimated $T_0$ is about 30 K, much larger than that of CeRu$_2$As$_2$, demonstrating stronger Kondo coupling. We should also mention that $C_{4f}/T$ displays a broad maximum near 40 K, see the right frame of the inset to Fig.~\ref{Fig5}(b). Such a broad peak arises from the Schottky anomaly which is a consequence of CEF splitting. The six-degenerated $j$=5/2 multiplet splits into three doublets in the presence of tetragonal CEF. The energy difference between the first excited and ground doublets is expected to be $\sim$90 K in this case, which agrees well with the hump observed in magnetic susceptibility [cf Fig.~\ref{Fig2}(b)].

These physical properties above enable us to place CeRu$_2$As$_2$ and CeIr$_2$As$_2$ respectively in the regime of magnetic ordered for small $J_{cf}$ and heavy-fermion for moderate $J_{cf}$. Aside from them, CeIr$_2$P$_2$ sits in the intermediate-valence regime with strong $J_{cf}$. Firstly, it is worthwhile to make a rough estimate of cerium valence through the bond-valence theory. The original idea was proposed by Brown\cite{Brown-BVP1981,Brown-BVP1985}, and the valence of an ion in the compound is a function of the bond lengths $d_{ij}$,
\begin{equation}
V_i=\sum_j\nu_{ij},
\label{Eq1}
\end{equation}
where $\nu_{ij}$ can be expressed in terms of $d_{ij}$\cite{Brese-BVP1991},
\begin{equation}
\nu_{ij}=\exp[(R_{ij}-d_{ij})/b].
\label{Eq2}
\end{equation}
Here $b$ is commonly taken to be a ``universal" constant close to 0.37 \AA, while $R_{ij}$ is called bond-valence parameter and is taken as 2.78 \AA~for Ce-As bonding and 2.70\AA~ for Ce-P bonding\cite{Brese-BVP1991}. According to the crystalline parameters, we get the cerium valences $+$2.27, $+$2.28 and $+$2.53 for CeRu$_2$As$_2$, CeIr$_2$As$_2$ and CeIr$_2$P$_2$, respectively. These values apparently are underestimated as compared to Ce$^{3+/4+}$, because we only take into account the Ce-As(P) bonding with nearest neighbour. However, what are important here are: (\rmnum{1}) the cerium valence in CeIr$_2$P$_2$ is much higher than in CeIr$_2$As$_2$, in line with the fact that CeIr$_2$P$_2$ has an intermediate valence; (\rmnum{2}) the calculated cerium valences are essentially the same in CeRu$_2$As$_2$ and CeIr$_2$As$_2$. This indicates that the transition metal is crucial to the physical properties of Ce$Tm_2$As$_2$. Indeed, $5d$ orbitals are more extended in space than $4d$ orbitals, and this makes the $c$-$f$ hybridization more effective in CeIr$_2$As$_2$. Apart from this, as we already mentioned, the CaBe$_2$Ge$_2$-type CeIr$_2$As$_2$ has the interlayer Ir-As bonding, and this results in a three-dimensional Ir-As network, and correspondingly, Kondo coupling is more efficient. A more straightforward case is CeNi$_2$As$_2$, which has both types of crystal structures\cite{Ghadraoui-LnNi2As2}. While the ThCr$_2$Si$_2$-type CeNi$_2$As$_2$ is an antiferromagnet with $T_N$$\approx$5 K\cite{LuoY-CeNi2As2}, the CaBe$_2$Ge$_2$-type CeNi$_2$As$_2$ is nonmagnetic due to Kondo effect\cite{Suzuki-CeNi2X2}. We argue that for these two reasons, the CaBe$_2$Ge$_2$-type CeIr$_2$As$_2$ exhibits much stronger Kondo effect and electronic correlation than the ThCr$_2$Si$_2$-type CeRu$_2$As$_2$.

\section{\Rmnum{4}. Conclusions}

To summarize, we investigated the physical properties of two Ce-122 Kondo lattice compounds, ThCr$_2$Si$_2$-type CeRu$_2$As$_2$ and CaBe$_2$Ge$_2$-type CeIr$_2$As$_2$. We find that CeRu$_2$As$_2$ is a local-moment antiferromagnet with N\'{e}el temperature $T_N$=4.3 K, Sommerfeld coefficient $\gamma_0$=35 mJ/mol$\cdot$K$^2$ and initial onset Kondo temperature $T_0$$\approx$6 K. The cerium moments are assumed to be uniaxially aligned along $\mathbf{c}$ which is also proposed as the easy-axis. CeIr$_2$As$_2$ appears to reside on the nonmagnetic side of a quantum critical point, exhibiting heavy-electron effect with enlarged $\gamma_0$$>$300 mJ/mol$\cdot$K$^2$ and $T_0$$\approx$30 K. This work, therefore, provides two new dense Kondo-lattice materials for further studying electronic correlation, quantum criticality and heavy-electron effects. High-quality single crystals are highly needed in the future.

\section{Acknowledgments}

The authors acknowledge Joe D. Thompson and Jinke Bao for helpful discussions. Y. Luo acknowledges 1000 Youth Talents Plan of China. Y. Li is supported by National Natural Science Foundation of China (Grant No. U1932155).

\setcounter{table}{0}
\setcounter{figure}{0}
\setcounter{equation}{0}
\renewcommand{\thefigure}{A\arabic{figure}}
\renewcommand{\thetable}{A\arabic{table}}
\renewcommand{\theequation}{A\arabic{equation}}
%\onecolumngrid

\section{Appendix: M\lowercase{agnetic susceptibility for aligned powders}}

\begin{figure}[!htp]
\vspace*{-80pt}
\hspace*{-18pt}
\includegraphics[width=9.7cm]{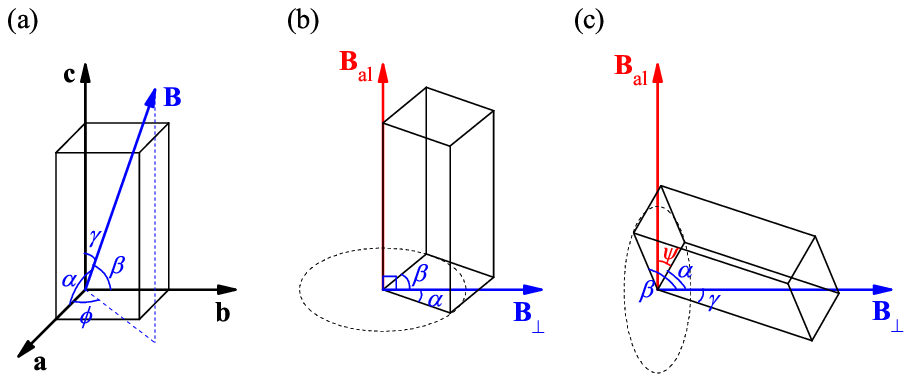}
\caption{\label{FigA1} (a) External magnetic field $\mathbf{B}$ applied in an arbitrary direction with respect to the \textit{tetragonal} unit cell. $\alpha$, $\beta$ and $\gamma$ are the angles between $\mathbf{B}$ and the principal axes, and $\phi$ is the angle spanned by $\mathbf{a}$ and the projection of $\mathbf{B}$ in $\mathbf{ab}$ plane. (b) For the easy-axis case, $\mathbf{c}$ is along the aligning field $\mathbf{B_{al}}$. (c) For the easy-plane case, $\mathbf{B_{al}}$ is inside $\mathbf{ab}$ plane, and $\mathbf{c}$ is perpendicular to $\mathbf{B_{al}}$. $\mathbf{B_{\perp}}$ ($\mathbf{B_{\parallel}}$, not shown) is the field applied perpendicular (parallel) to $\mathbf{B_{al}}$ to measure $m_{\perp}$ ($m_{\parallel}$).}
\end{figure}

To start with, we deduce the formulas of magnetic susceptibility of polycrystalline powders. Assuming an arbitrary magnetic field $\mathbf{B}$ is applied to a tetragonal unit cell (e.g. CeRu$_2$As$_2$), the direction of field is characterized by $\alpha$, $\beta$ and $\gamma$ as shown in Fig.~\ref{FigA1}(a). The magnetic susceptibility in such configuration is $\chi(\alpha,\beta,\gamma)$=$\chi_{a}\cos^2\alpha$$+$$\chi_{b}\cos^2\beta$$+$$\chi_{c}\cos^2\gamma$ \cite{Boutron-PMchi}. Note that $\chi_a$=$\chi_b$ for tetragonal symmetry. Taking the powder average, we derive the susceptibility for a bulk polycrystal
\begin{equation}
\chi_{\text{pow}}=\frac{\int\int \chi(\alpha,\beta,\gamma)\sin\gamma d\gamma d\phi}{\int_0^{2\pi} \int_0^{\pi} \sin\gamma d\gamma d\phi}=\frac{\chi_a+\chi_b+\chi_c}{3},
\label{EqA1}
\end{equation}
where $\phi$ is the angle between $\mathbf{a}$ and the projection of $\mathbf{B}$ in $\mathbf{ab}$ plane.

Now considering that the powder is aligned by an aligning field $\mathbf{B_{al}}$:\\
$\bullet$ For an easy-axis case [Fig.~\ref{FigA1}(b)], $\mathbf{c}$ will be aligned to $\mathbf{B_{al}}$, and one easily finds $\chi_{\parallel}$=$\chi_c$, while $\chi_{\perp}$=$\chi_{a,b}$. Here, the notations $\parallel$ and $\perp$ are with respect to $\mathbf{B_{al}}$.\\
$\bullet$ For an easy-plane case [Fig.~\ref{FigA1}(c)], $\mathbf{ab}$ plane will be parallel with $\mathbf{B_{al}}$, while $\mathbf{c}$ lays inside the plane perpendicular to $\mathbf{B_{al}}$. The angle between $\mathbf{a}$ and $\mathbf{B_{al}}$ is named $\psi$. In this case, $\chi_{\parallel}$=$\chi_{a,b}$, while $\chi_{\perp}$ is
\begin{equation}
\chi_{\perp}=\frac{\int\int \chi(\alpha,\beta,\gamma)\sin\psi d\psi d\gamma}{\int_0^{2\pi}\int_0^{\pi} \sin\psi d\psi d\gamma}=\frac{\chi_{a,b}+\chi_c}{2}.
\label{EqA2}
\end{equation}
Eq.~(\ref{EqA2}) can be deduced, because $\cos\alpha$=$\sin\psi\sin\gamma$, and $\cos\beta$=$-\cos\psi\sin\gamma$. Interestingly, for both easy-axis and easy-plane cases,
\begin{equation}
\chi_{\parallel}+2\chi_{\perp}=\chi_a+\chi_b+\chi_c=3\chi_{\text{pow}}.
\label{EqA3}
\end{equation}

\begin{figure}[!htp]
\vspace{-5pt}
\hspace{-10pt}
\includegraphics[width=9.0cm]{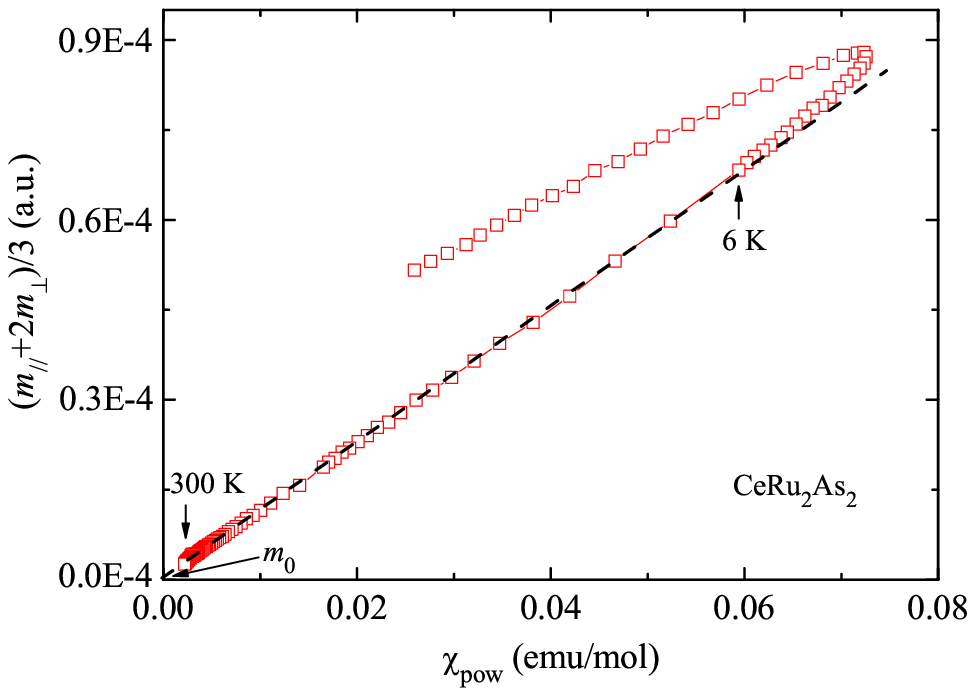}
\vspace{-25pt}
\caption{\label{FigA2} ($m_{\parallel}$+2$m_{\perp})/3$ vs. $\chi_{\text{pow}}$ with $T$ as an implicit parameter. The slope is determined by the weight of aligned CeRu$_2$As$_2$ powders.}
\end{figure}

Eq.~(\ref{EqA3}) enables us to figure out the weight of aligned powers in the mixture. In Fig.~\ref{FigA2} we plot $(m_{\parallel}+2m_{\perp})/3$ vs. $\chi_{\text{pow}}$ where $T$ is the implicit parameter. The plot shows good linearity for $T$ above 6 K. The intercept of this linear scaling, $m_0$, should be attributed to Stycast (which we assume is isotropic and temperature independent), and the slope is determined by the weight of the aligned powders. After subtracting $m_0$ from $m_{\parallel}$ and $m_{\perp}$, and with the slope, we are able to convert magnetization into magnetic susceptibility, as shown in Fig.~\ref{Fig3}(a).

%\bibliography{biblio}

\begin{thebibliography}{43}%
\makeatletter
\providecommand \@ifxundefined [1]{%
 \@ifx{#1\undefined}
}%
\providecommand \@ifnum [1]{%
 \ifnum #1\expandafter \@firstoftwo
 \else \expandafter \@secondoftwo
 \fi
}%
\providecommand \@ifx [1]{%
 \ifx #1\expandafter \@firstoftwo
 \else \expandafter \@secondoftwo
 \fi
}%
\providecommand \natexlab [1]{#1}%
\providecommand \enquote  [1]{``#1''}%
\providecommand \bibnamefont  [1]{#1}%
\providecommand \bibfnamefont [1]{#1}%
\providecommand \citenamefont [1]{#1}%
\providecommand \href@noop [0]{\@secondoftwo}%
\providecommand \href [0]{\begingroup \@sanitize@url \@href}%
\providecommand \@href[1]{\@@startlink{#1}\@@href}%
\providecommand \@@href[1]{\endgroup#1\@@endlink}%
\providecommand \@sanitize@url [0]{\catcode `\\12\catcode `\$12\catcode
  `\&12\catcode `\#12\catcode `\^12\catcode `\_12\catcode `\%12\relax}%
\providecommand \@@startlink[1]{}%
\providecommand \@@endlink[0]{}%
\providecommand \url  [0]{\begingroup\@sanitize@url \@url }%
\providecommand \@url [1]{\endgroup\@href {#1}{\urlprefix }}%
\providecommand \urlprefix  [0]{URL }%
\providecommand \Eprint [0]{\href }%
\providecommand \doibase [0]{http://dx.doi.org/}%
\providecommand \selectlanguage [0]{\@gobble}%
\providecommand \bibinfo  [0]{\@secondoftwo}%
\providecommand \bibfield  [0]{\@secondoftwo}%
\providecommand \translation [1]{[#1]}%
\providecommand \BibitemOpen [0]{}%
\providecommand \bibitemStop [0]{}%
\providecommand \bibitemNoStop [0]{.\EOS\space}%
\providecommand \EOS [0]{\spacefactor3000\relax}%
\providecommand \BibitemShut  [1]{\csname bibitem#1\endcsname}%
\let\auto@bib@innerbib\@empty
%</preamble>
\bibitem [{\citenamefont {Ruderman}\ and\ \citenamefont
  {Kittel}(1954)}]{RKKY-RK}%
  \BibitemOpen
  \bibfield  {author} {\bibinfo {author} {\bibfnamefont {M.~A.}\ \bibnamefont
  {Ruderman}}\ and\ \bibinfo {author} {\bibfnamefont {C.}~\bibnamefont
  {Kittel}},\ }\href {\doibase 10.1103/PhysRev.96.99} {\bibfield  {journal}
  {\bibinfo  {journal} {Phys. Rev.}\ }\textbf {\bibinfo {volume} {96}},\
  \bibinfo {pages} {99} (\bibinfo {year} {1954})}\BibitemShut {NoStop}%
\bibitem [{\citenamefont {Kasuya}(1956)}]{RKKY-K}%
  \BibitemOpen
  \bibfield  {author} {\bibinfo {author} {\bibfnamefont {T.}~\bibnamefont
  {Kasuya}},\ }\href {\doibase 10.1143/PTP.16.45} {\bibfield  {journal}
  {\bibinfo  {journal} {Prog. Theor. Phys.}\ }\textbf {\bibinfo {volume}
  {16}},\ \bibinfo {pages} {45} (\bibinfo {year} {1956})}\BibitemShut {NoStop}%
\bibitem [{\citenamefont {Yosida}(1957)}]{RKKY-Y}%
  \BibitemOpen
  \bibfield  {author} {\bibinfo {author} {\bibfnamefont {K.}~\bibnamefont
  {Yosida}},\ }\href {\doibase 10.1103/PhysRev.106.893} {\bibfield  {journal}
  {\bibinfo  {journal} {Phys. Rev.}\ }\textbf {\bibinfo {volume} {106}},\
  \bibinfo {pages} {893} (\bibinfo {year} {1957})}\BibitemShut {NoStop}%
\bibitem [{\citenamefont {Kondo}(1964)}]{Kondo-RMinimum}%
  \BibitemOpen
  \bibfield  {author} {\bibinfo {author} {\bibfnamefont {J.}~\bibnamefont
  {Kondo}},\ }\href {\doibase 10.1143/PTP.32.37} {\bibfield  {journal}
  {\bibinfo  {journal} {Prog. Theor. Phys}\ }\textbf {\bibinfo {volume} {32}},\
  \bibinfo {pages} {37} (\bibinfo {year} {1964})}\BibitemShut {NoStop}%
\bibitem [{\citenamefont {Hewson}(1993)}]{Hewson-Kondo}%
  \BibitemOpen
  \bibfield  {author} {\bibinfo {author} {\bibfnamefont {A.~C.}\ \bibnamefont
  {Hewson}},\ }\href@noop {} {\emph {\bibinfo {title} {{The Kondo Problem to
  Heavy Fermions}}}}\ (\bibinfo  {publisher} {Cambridge University Press,
  Cambridge},\ \bibinfo {year} {1993})\BibitemShut {NoStop}%
\bibitem [{\citenamefont {Rotter}\ \emph {et~al.}(2008)\citenamefont {Rotter},
  \citenamefont {Tegel},\ and\ \citenamefont {Johrendt}}]{Rotter-BaK122SC}%
  \BibitemOpen
  \bibfield  {author} {\bibinfo {author} {\bibfnamefont {M.}~\bibnamefont
  {Rotter}}, \bibinfo {author} {\bibfnamefont {M.}~\bibnamefont {Tegel}}, \
  and\ \bibinfo {author} {\bibfnamefont {D.}~\bibnamefont {Johrendt}},\ }\href
  {\doibase 10.1103/PhysRevLett.101.107006} {\bibfield  {journal} {\bibinfo
  {journal} {Phys. Rev. Lett.}\ }\textbf {\bibinfo {volume} {101}},\ \bibinfo
  {pages} {107006} (\bibinfo {year} {2008})}\BibitemShut {NoStop}%
\bibitem [{\citenamefont {Stewart}(2001)}]{Stewart-RMPNFL}%
  \BibitemOpen
  \bibfield  {author} {\bibinfo {author} {\bibfnamefont {G.~R.}\ \bibnamefont
  {Stewart}},\ }\href {\doibase 10.1103/RevModPhys.73.797} {\bibfield
  {journal} {\bibinfo  {journal} {Rev. Mod. Phys.}\ }\textbf {\bibinfo {volume}
  {73}},\ \bibinfo {pages} {797} (\bibinfo {year} {2001})}\BibitemShut
  {NoStop}%
\bibitem [{\citenamefont {Misra}(2008)}]{Misra-HFSystem}%
  \BibitemOpen
  \bibfield  {author} {\bibinfo {author} {\bibfnamefont {P.}~\bibnamefont
  {Misra}},\ }\href@noop {} {\emph {\bibinfo {title} {{Handbook of Metal
  Physics: Heavy-Fermion Systems}}}}\ (\bibinfo  {publisher} {Elsevier,
  Amsterdam},\ \bibinfo {year} {2008})\BibitemShut {NoStop}%
\bibitem [{\citenamefont {Pfleiderer}(2009)}]{Pfleiderer-RMPHFSC}%
  \BibitemOpen
  \bibfield  {author} {\bibinfo {author} {\bibfnamefont {C.}~\bibnamefont
  {Pfleiderer}},\ }\href {\doibase 10.1103/RevModPhys.81.1551} {\bibfield
  {journal} {\bibinfo  {journal} {Rev. Mod. Phys.}\ }\textbf {\bibinfo {volume}
  {81}},\ \bibinfo {pages} {1551} (\bibinfo {year} {2009})}\BibitemShut
  {NoStop}%
\bibitem [{\citenamefont {Steglich}\ \emph {et~al.}(1979)\citenamefont
  {Steglich}, \citenamefont {Aarts}, \citenamefont {Bredl}, \citenamefont
  {Lieke}, \citenamefont {Meschede}, \citenamefont {Franz},\ and\ \citenamefont
  {Sch\"afer}}]{Steglich-CeCu2Si2SC}%
  \BibitemOpen
  \bibfield  {author} {\bibinfo {author} {\bibfnamefont {F.}~\bibnamefont
  {Steglich}}, \bibinfo {author} {\bibfnamefont {J.}~\bibnamefont {Aarts}},
  \bibinfo {author} {\bibfnamefont {C.~D.}\ \bibnamefont {Bredl}}, \bibinfo
  {author} {\bibfnamefont {W.}~\bibnamefont {Lieke}}, \bibinfo {author}
  {\bibfnamefont {D.}~\bibnamefont {Meschede}}, \bibinfo {author}
  {\bibfnamefont {W.}~\bibnamefont {Franz}}, \ and\ \bibinfo {author}
  {\bibfnamefont {H.}~\bibnamefont {Sch\"afer}},\ }\href {\doibase
  10.1103/PhysRevLett.43.1892} {\bibfield  {journal} {\bibinfo  {journal}
  {Phys. Rev. Lett.}\ }\textbf {\bibinfo {volume} {43}},\ \bibinfo {pages}
  {1892} (\bibinfo {year} {1979})}\BibitemShut {NoStop}%
\bibitem [{\citenamefont {Palstra}\ \emph {et~al.}(1985)\citenamefont
  {Palstra}, \citenamefont {Menovsky}, \citenamefont {Berg}, \citenamefont
  {Dirkmaat}, \citenamefont {Kes}, \citenamefont {Nieuwenhuys},\ and\
  \citenamefont {Mydosh}}]{Palstra-URu2Si2HO}%
  \BibitemOpen
  \bibfield  {author} {\bibinfo {author} {\bibfnamefont {T.~T.~M.}\
  \bibnamefont {Palstra}}, \bibinfo {author} {\bibfnamefont {A.~A.}\
  \bibnamefont {Menovsky}}, \bibinfo {author} {\bibfnamefont {J.~v.~d.}\
  \bibnamefont {Berg}}, \bibinfo {author} {\bibfnamefont {A.~J.}\ \bibnamefont
  {Dirkmaat}}, \bibinfo {author} {\bibfnamefont {P.~H.}\ \bibnamefont {Kes}},
  \bibinfo {author} {\bibfnamefont {G.~J.}\ \bibnamefont {Nieuwenhuys}}, \ and\
  \bibinfo {author} {\bibfnamefont {J.~A.}\ \bibnamefont {Mydosh}},\ }\href
  {\doibase 10.1103/PhysRevLett.55.2727} {\bibfield  {journal} {\bibinfo
  {journal} {Phys. Rev. Lett.}\ }\textbf {\bibinfo {volume} {55}},\ \bibinfo
  {pages} {2727} (\bibinfo {year} {1985})}\BibitemShut {NoStop}%
\bibitem [{\citenamefont {Custers}\ \emph {et~al.}(2003)\citenamefont
  {Custers}, \citenamefont {Gegenwart}, \citenamefont {Wilhelm}, \citenamefont
  {Neumaier}, \citenamefont {Tokiwa}, \citenamefont {Trovarelli}, \citenamefont
  {Geibel}, \citenamefont {Steglich}, \citenamefont {Pepin},\ and\
  \citenamefont {Coleman}}]{Custers-YbRh2Si2QCP}%
  \BibitemOpen
  \bibfield  {author} {\bibinfo {author} {\bibfnamefont {J.}~\bibnamefont
  {Custers}}, \bibinfo {author} {\bibfnamefont {P.}~\bibnamefont {Gegenwart}},
  \bibinfo {author} {\bibfnamefont {H.}~\bibnamefont {Wilhelm}}, \bibinfo
  {author} {\bibfnamefont {K.}~\bibnamefont {Neumaier}}, \bibinfo {author}
  {\bibfnamefont {Y.}~\bibnamefont {Tokiwa}}, \bibinfo {author} {\bibfnamefont
  {O.}~\bibnamefont {Trovarelli}}, \bibinfo {author} {\bibfnamefont
  {C.}~\bibnamefont {Geibel}}, \bibinfo {author} {\bibfnamefont
  {F.}~\bibnamefont {Steglich}}, \bibinfo {author} {\bibfnamefont
  {C.}~\bibnamefont {Pepin}}, \ and\ \bibinfo {author} {\bibfnamefont
  {P.}~\bibnamefont {Coleman}},\ }\href {\doibase 10.1038/nature01774}
  {\bibfield  {journal} {\bibinfo  {journal} {Nature}\ }\textbf {\bibinfo
  {volume} {424}},\ \bibinfo {pages} {524} (\bibinfo {year}
  {2003})}\BibitemShut {NoStop}%
\bibitem [{\citenamefont {Madar}\ \emph {et~al.}(1987)\citenamefont {Madar},
  \citenamefont {Chaudouet}, \citenamefont {Senateur}, \citenamefont {Zemni},\
  and\ \citenamefont {Tranqui}}]{Madar-CeRh2P2}%
  \BibitemOpen
  \bibfield  {author} {\bibinfo {author} {\bibfnamefont {R.}~\bibnamefont
  {Madar}}, \bibinfo {author} {\bibfnamefont {P.}~\bibnamefont {Chaudouet}},
  \bibinfo {author} {\bibfnamefont {J.}~\bibnamefont {Senateur}}, \bibinfo
  {author} {\bibfnamefont {S.}~\bibnamefont {Zemni}}, \ and\ \bibinfo {author}
  {\bibfnamefont {D.}~\bibnamefont {Tranqui}},\ }\href {\doibase
  10.1016/0022-5088(87)90241-4} {\bibfield  {journal} {\bibinfo  {journal} {J.
  Less Common Met.}\ }\textbf {\bibinfo {volume} {133}},\ \bibinfo {pages}
  {303} (\bibinfo {year} {1987})}\BibitemShut {NoStop}%
\bibitem [{\citenamefont {Ghadraoui}\ \emph {et~al.}(1988)\citenamefont
  {Ghadraoui}, \citenamefont {Pivan}, \citenamefont {Gu\'{e}rin}, \citenamefont
  {Pena}, \citenamefont {Padiou},\ and\ \citenamefont
  {Sergent}}]{Ghadraoui-LnNi2As2}%
  \BibitemOpen
  \bibfield  {author} {\bibinfo {author} {\bibfnamefont {E.~E.}\ \bibnamefont
  {Ghadraoui}}, \bibinfo {author} {\bibfnamefont {J.}~\bibnamefont {Pivan}},
  \bibinfo {author} {\bibfnamefont {R.}~\bibnamefont {Gu\'{e}rin}}, \bibinfo
  {author} {\bibfnamefont {O.}~\bibnamefont {Pena}}, \bibinfo {author}
  {\bibfnamefont {J.}~\bibnamefont {Padiou}}, \ and\ \bibinfo {author}
  {\bibfnamefont {M.}~\bibnamefont {Sergent}},\ }\href {\doibase
  10.1016/0025-5408(88)90123-7} {\bibfield  {journal} {\bibinfo  {journal}
  {Mater. Res. Bull.}\ }\textbf {\bibinfo {volume} {23}},\ \bibinfo {pages}
  {1345} (\bibinfo {year} {1988})}\BibitemShut {NoStop}%
\bibitem [{\citenamefont {Hiebl}\ \emph {et~al.}(1986)\citenamefont {Hiebl},
  \citenamefont {Horvath},\ and\ \citenamefont {Rogl}}]{Hiebl-CeT2Si2}%
  \BibitemOpen
  \bibfield  {author} {\bibinfo {author} {\bibfnamefont {K.}~\bibnamefont
  {Hiebl}}, \bibinfo {author} {\bibfnamefont {C.}~\bibnamefont {Horvath}}, \
  and\ \bibinfo {author} {\bibfnamefont {P.}~\bibnamefont {Rogl}},\ }\href
  {\doibase 10.1016/0022-5088(86)90063-9} {\bibfield  {journal} {\bibinfo
  {journal} {J. Less Common Met.}\ }\textbf {\bibinfo {volume} {117}},\
  \bibinfo {pages} {375} (\bibinfo {year} {1986})}\BibitemShut {NoStop}%
\bibitem [{\citenamefont {Suzuki}\ \emph {et~al.}(2001)\citenamefont {Suzuki},
  \citenamefont {Abe}, \citenamefont {Kitazawa},\ and\ \citenamefont
  {Schmitt}}]{Suzuki-CeNi2X2}%
  \BibitemOpen
  \bibfield  {author} {\bibinfo {author} {\bibfnamefont {H.}~\bibnamefont
  {Suzuki}}, \bibinfo {author} {\bibfnamefont {H.}~\bibnamefont {Abe}},
  \bibinfo {author} {\bibfnamefont {H.}~\bibnamefont {Kitazawa}}, \ and\
  \bibinfo {author} {\bibfnamefont {D.}~\bibnamefont {Schmitt}},\ }\href
  {\doibase 10.1016/S0925-8388(01)01164-1} {\bibfield  {journal} {\bibinfo
  {journal} {J. Alloys Comp.}\ }\textbf {\bibinfo {volume} {323-324}},\
  \bibinfo {pages} {520} (\bibinfo {year} {2001})}\BibitemShut {NoStop}%
\bibitem [{\citenamefont {Luo}\ \emph {et~al.}(2012)\citenamefont {Luo},
  \citenamefont {Bao}, \citenamefont {Shen}, \citenamefont {Han}, \citenamefont
  {Yang}, \citenamefont {Lv}, \citenamefont {Li}, \citenamefont {Jiao},
  \citenamefont {Si}, \citenamefont {Feng}, \citenamefont {Dai}, \citenamefont
  {Cao},\ and\ \citenamefont {Xu}}]{LuoY-CeNi2As2}%
  \BibitemOpen
  \bibfield  {author} {\bibinfo {author} {\bibfnamefont {Y.}~\bibnamefont
  {Luo}}, \bibinfo {author} {\bibfnamefont {J.}~\bibnamefont {Bao}}, \bibinfo
  {author} {\bibfnamefont {C.}~\bibnamefont {Shen}}, \bibinfo {author}
  {\bibfnamefont {J.}~\bibnamefont {Han}}, \bibinfo {author} {\bibfnamefont
  {X.}~\bibnamefont {Yang}}, \bibinfo {author} {\bibfnamefont {C.}~\bibnamefont
  {Lv}}, \bibinfo {author} {\bibfnamefont {Y.}~\bibnamefont {Li}}, \bibinfo
  {author} {\bibfnamefont {W.}~\bibnamefont {Jiao}}, \bibinfo {author}
  {\bibfnamefont {B.}~\bibnamefont {Si}}, \bibinfo {author} {\bibfnamefont
  {C.}~\bibnamefont {Feng}}, \bibinfo {author} {\bibfnamefont {J.}~\bibnamefont
  {Dai}}, \bibinfo {author} {\bibfnamefont {G.}~\bibnamefont {Cao}}, \ and\
  \bibinfo {author} {\bibfnamefont {Z.-a.}\ \bibnamefont {Xu}},\ }\href
  {\doibase 10.1103/PhysRevB.86.245130} {\bibfield  {journal} {\bibinfo
  {journal} {Phys. Rev. B}\ }\textbf {\bibinfo {volume} {86}},\ \bibinfo
  {pages} {245130} (\bibinfo {year} {2012})}\BibitemShut {NoStop}%
\bibitem [{\citenamefont {Flouquet}\ \emph {et~al.}(2002)\citenamefont
  {Flouquet}, \citenamefont {Haen}, \citenamefont {Raymond}, \citenamefont
  {Aoki},\ and\ \citenamefont {Knebel}}]{Flouquet-CeRu2Si2MM}%
  \BibitemOpen
  \bibfield  {author} {\bibinfo {author} {\bibfnamefont {J.}~\bibnamefont
  {Flouquet}}, \bibinfo {author} {\bibfnamefont {P.}~\bibnamefont {Haen}},
  \bibinfo {author} {\bibfnamefont {S.}~\bibnamefont {Raymond}}, \bibinfo
  {author} {\bibfnamefont {D.}~\bibnamefont {Aoki}}, \ and\ \bibinfo {author}
  {\bibfnamefont {G.}~\bibnamefont {Knebel}},\ }\href {\doibase
  10.1016/S0921-4526(02)01126-2} {\bibfield  {journal} {\bibinfo  {journal}
  {Physica B}\ }\textbf {\bibinfo {volume} {319}},\ \bibinfo {pages} {251}
  (\bibinfo {year} {2002})}\BibitemShut {NoStop}%
\bibitem [{\citenamefont {Yuan}\ \emph {et~al.}(2003)\citenamefont {Yuan},
  \citenamefont {Grosche}, \citenamefont {Deppe}, \citenamefont {Geibel},
  \citenamefont {Sparn},\ and\ \citenamefont {Steglich}}]{Yuan-CeCu2Si2TwoSC}%
  \BibitemOpen
  \bibfield  {author} {\bibinfo {author} {\bibfnamefont {H.~Q.}\ \bibnamefont
  {Yuan}}, \bibinfo {author} {\bibfnamefont {F.~M.}\ \bibnamefont {Grosche}},
  \bibinfo {author} {\bibfnamefont {M.}~\bibnamefont {Deppe}}, \bibinfo
  {author} {\bibfnamefont {C.}~\bibnamefont {Geibel}}, \bibinfo {author}
  {\bibfnamefont {G.}~\bibnamefont {Sparn}}, \ and\ \bibinfo {author}
  {\bibfnamefont {F.}~\bibnamefont {Steglich}},\ }\href {\doibase
  10.1126/science.1091648} {\bibfield  {journal} {\bibinfo  {journal}
  {Science}\ }\textbf {\bibinfo {volume} {302}},\ \bibinfo {pages} {2104}
  (\bibinfo {year} {2003})}\BibitemShut {NoStop}%
\bibitem [{\citenamefont {Grosche}\ \emph {et~al.}(1996)\citenamefont
  {Grosche}, \citenamefont {Julian}, \citenamefont {Mathur},\ and\
  \citenamefont {Lonzarich}}]{Grosche-CePd2Si2SC}%
  \BibitemOpen
  \bibfield  {author} {\bibinfo {author} {\bibfnamefont {F.}~\bibnamefont
  {Grosche}}, \bibinfo {author} {\bibfnamefont {S.}~\bibnamefont {Julian}},
  \bibinfo {author} {\bibfnamefont {N.}~\bibnamefont {Mathur}}, \ and\ \bibinfo
  {author} {\bibfnamefont {G.}~\bibnamefont {Lonzarich}},\ }\href {\doibase
  10.1016/0921-4526(96)00036-1} {\bibfield  {journal} {\bibinfo  {journal}
  {Physica B}\ }\textbf {\bibinfo {volume} {223-224}},\ \bibinfo {pages} {50}
  (\bibinfo {year} {1996})}\BibitemShut {NoStop}%
\bibitem [{\citenamefont {Movshovich}\ \emph {et~al.}(1996)\citenamefont
  {Movshovich}, \citenamefont {Graf}, \citenamefont {Mandrus}, \citenamefont
  {Thompson}, \citenamefont {Smith},\ and\ \citenamefont
  {Fisk}}]{Movshovich-CeRh2Si2SC}%
  \BibitemOpen
  \bibfield  {author} {\bibinfo {author} {\bibfnamefont {R.}~\bibnamefont
  {Movshovich}}, \bibinfo {author} {\bibfnamefont {T.}~\bibnamefont {Graf}},
  \bibinfo {author} {\bibfnamefont {D.}~\bibnamefont {Mandrus}}, \bibinfo
  {author} {\bibfnamefont {J.~D.}\ \bibnamefont {Thompson}}, \bibinfo {author}
  {\bibfnamefont {J.~L.}\ \bibnamefont {Smith}}, \ and\ \bibinfo {author}
  {\bibfnamefont {Z.}~\bibnamefont {Fisk}},\ }\href {\doibase
  10.1103/PhysRevB.53.8241} {\bibfield  {journal} {\bibinfo  {journal} {Phys.
  Rev. B}\ }\textbf {\bibinfo {volume} {53}},\ \bibinfo {pages} {8241}
  (\bibinfo {year} {1996})}\BibitemShut {NoStop}%
\bibitem [{\citenamefont {Luo}\ \emph {et~al.}(2015)\citenamefont {Luo},
  \citenamefont {Ronning}, \citenamefont {Wakeham}, \citenamefont {Lu},
  \citenamefont {Park}, \citenamefont {Xu},\ and\ \citenamefont
  {Thompson}}]{LuoY-CeNi2As2Pre}%
  \BibitemOpen
  \bibfield  {author} {\bibinfo {author} {\bibfnamefont {Y.}~\bibnamefont
  {Luo}}, \bibinfo {author} {\bibfnamefont {F.}~\bibnamefont {Ronning}},
  \bibinfo {author} {\bibfnamefont {N.}~\bibnamefont {Wakeham}}, \bibinfo
  {author} {\bibfnamefont {X.}~\bibnamefont {Lu}}, \bibinfo {author}
  {\bibfnamefont {T.}~\bibnamefont {Park}}, \bibinfo {author} {\bibfnamefont
  {Z.~A.}\ \bibnamefont {Xu}}, \ and\ \bibinfo {author} {\bibfnamefont {J.~D.}\
  \bibnamefont {Thompson}},\ }\href {\doibase 10.1073/pnas.1509581112}
  {\bibfield  {journal} {\bibinfo  {journal} {Proc. Natl. Acad. Sci. USA}\
  }\textbf {\bibinfo {volume} {112}},\ \bibinfo {pages} {13520} (\bibinfo
  {year} {2015})}\BibitemShut {NoStop}%
\bibitem [{\citenamefont {Pfannenschmidt}\ \emph {et~al.}(2012)\citenamefont
  {Pfannenschmidt}, \citenamefont {Behrends}, \citenamefont {Lincke},
  \citenamefont {Eul}, \citenamefont {Sch\"{a}fer}, \citenamefont {Eckert},\
  and\ \citenamefont {P\"{o}ttgen}}]{Pfannenschmidt-ReIr2Pn2}%
  \BibitemOpen
  \bibfield  {author} {\bibinfo {author} {\bibfnamefont {U.}~\bibnamefont
  {Pfannenschmidt}}, \bibinfo {author} {\bibfnamefont {F.}~\bibnamefont
  {Behrends}}, \bibinfo {author} {\bibfnamefont {H.}~\bibnamefont {Lincke}},
  \bibinfo {author} {\bibfnamefont {M.}~\bibnamefont {Eul}}, \bibinfo {author}
  {\bibfnamefont {K.}~\bibnamefont {Sch\"{a}fer}}, \bibinfo {author}
  {\bibfnamefont {H.}~\bibnamefont {Eckert}}, \ and\ \bibinfo {author}
  {\bibfnamefont {R.}~\bibnamefont {P\"{o}ttgen}},\ }\href {\doibase
  10.1039/C2DT31874A} {\bibfield  {journal} {\bibinfo  {journal} {Dalton
  Trans.}\ }\textbf {\bibinfo {volume} {41}},\ \bibinfo {pages} {14188}
  (\bibinfo {year} {2012})}\BibitemShut {NoStop}%
\bibitem [{\citenamefont {Petrovic}\ \emph
  {et~al.}(2001{\natexlab{a}})\citenamefont {Petrovic}, \citenamefont
  {Pagliuso}, \citenamefont {Hundley}, \citenamefont {Movshovich},
  \citenamefont {Sarrao}, \citenamefont {Thompson}, \citenamefont {Fisk},\ and\
  \citenamefont {Monthoux}}]{Petrovic-CeCoIn5SC}%
  \BibitemOpen
  \bibfield  {author} {\bibinfo {author} {\bibfnamefont {C.}~\bibnamefont
  {Petrovic}}, \bibinfo {author} {\bibfnamefont {P.~G.}\ \bibnamefont
  {Pagliuso}}, \bibinfo {author} {\bibfnamefont {M.~F.}\ \bibnamefont
  {Hundley}}, \bibinfo {author} {\bibfnamefont {R.}~\bibnamefont {Movshovich}},
  \bibinfo {author} {\bibfnamefont {J.~L.}\ \bibnamefont {Sarrao}}, \bibinfo
  {author} {\bibfnamefont {J.~D.}\ \bibnamefont {Thompson}}, \bibinfo {author}
  {\bibfnamefont {Z.}~\bibnamefont {Fisk}}, \ and\ \bibinfo {author}
  {\bibfnamefont {P.}~\bibnamefont {Monthoux}},\ }\href {\doibase
  10.1088/0953-8984/13/17/103} {\bibfield  {journal} {\bibinfo  {journal} {J.
  Phys.: Condens. Matter}\ }\textbf {\bibinfo {volume} {13}},\ \bibinfo {pages}
  {L337} (\bibinfo {year} {2001}{\natexlab{a}})}\BibitemShut {NoStop}%
\bibitem [{\citenamefont {Mun}\ \emph {et~al.}(2010)\citenamefont {Mun},
  \citenamefont {Bud'ko}, \citenamefont {Kreyssig},\ and\ \citenamefont
  {Canfield}}]{Mun-CeGeNi3}%
  \BibitemOpen
  \bibfield  {author} {\bibinfo {author} {\bibfnamefont {E.~D.}\ \bibnamefont
  {Mun}}, \bibinfo {author} {\bibfnamefont {S.~L.}\ \bibnamefont {Bud'ko}},
  \bibinfo {author} {\bibfnamefont {A.}~\bibnamefont {Kreyssig}}, \ and\
  \bibinfo {author} {\bibfnamefont {P.~C.}\ \bibnamefont {Canfield}},\ }\href
  {\doibase 10.1103/PhysRevB.82.054424} {\bibfield  {journal} {\bibinfo
  {journal} {Phys. Rev. B}\ }\textbf {\bibinfo {volume} {82}},\ \bibinfo
  {pages} {054424} (\bibinfo {year} {2010})}\BibitemShut {NoStop}%
\bibitem [{\citenamefont {Blundell}(2001)}]{Blundell-MaginCM}%
  \BibitemOpen
  \bibfield  {author} {\bibinfo {author} {\bibfnamefont {S.}~\bibnamefont
  {Blundell}},\ }\href@noop {} {\emph {\bibinfo {title} {{Magnetism in
  Condensed Matter}}}}\ (\bibinfo  {publisher} {Oxford University Press},\
  \bibinfo {year} {2001})\BibitemShut {NoStop}%
\bibitem [{\citenamefont {Young}\ \emph {et~al.}(2002)\citenamefont {Young},
  \citenamefont {Rose}, \citenamefont {MacLaughlin}, \citenamefont {Ishida},
  \citenamefont {Bernal}, \citenamefont {Lukefahr}, \citenamefont {Heuser},
  \citenamefont {Freeman},\ and\ \citenamefont {Maple}}]{Young-Alignment}%
  \BibitemOpen
  \bibfield  {author} {\bibinfo {author} {\bibfnamefont {B.-L.}\ \bibnamefont
  {Young}}, \bibinfo {author} {\bibfnamefont {M.~S.}\ \bibnamefont {Rose}},
  \bibinfo {author} {\bibfnamefont {D.~E.}\ \bibnamefont {MacLaughlin}},
  \bibinfo {author} {\bibfnamefont {K.}~\bibnamefont {Ishida}}, \bibinfo
  {author} {\bibfnamefont {O.~O.}\ \bibnamefont {Bernal}}, \bibinfo {author}
  {\bibfnamefont {H.~G.}\ \bibnamefont {Lukefahr}}, \bibinfo {author}
  {\bibfnamefont {K.}~\bibnamefont {Heuser}}, \bibinfo {author} {\bibfnamefont
  {E.~J.}\ \bibnamefont {Freeman}}, \ and\ \bibinfo {author} {\bibfnamefont
  {M.~B.}\ \bibnamefont {Maple}},\ }\href {\doibase 10.1063/1.1488682}
  {\bibfield  {journal} {\bibinfo  {journal} {Rev. Sci. Instrum.}\ }\textbf
  {\bibinfo {volume} {73}},\ \bibinfo {pages} {3038} (\bibinfo {year}
  {2002})}\BibitemShut {NoStop}%
\bibitem [{\citenamefont {Guo}\ \emph {et~al.}(2016)\citenamefont {Guo},
  \citenamefont {Pan}, \citenamefont {Yu}, \citenamefont {Ruan}, \citenamefont
  {Chen}, \citenamefont {Wang}, \citenamefont {Mu}, \citenamefont {Chen},\ and\
  \citenamefont {Ren}}]{Guo-LaRu2As2SC}%
  \BibitemOpen
  \bibfield  {author} {\bibinfo {author} {\bibfnamefont {Q.}~\bibnamefont
  {Guo}}, \bibinfo {author} {\bibfnamefont {B.-J.}\ \bibnamefont {Pan}},
  \bibinfo {author} {\bibfnamefont {J.}~\bibnamefont {Yu}}, \bibinfo {author}
  {\bibfnamefont {B.-B.}\ \bibnamefont {Ruan}}, \bibinfo {author}
  {\bibfnamefont {D.-Y.}\ \bibnamefont {Chen}}, \bibinfo {author}
  {\bibfnamefont {X.-C.}\ \bibnamefont {Wang}}, \bibinfo {author}
  {\bibfnamefont {Q.-G.}\ \bibnamefont {Mu}}, \bibinfo {author} {\bibfnamefont
  {G.-F.}\ \bibnamefont {Chen}}, \ and\ \bibinfo {author} {\bibfnamefont
  {Z.-A.}\ \bibnamefont {Ren}},\ }\href {\doibase 10.1007/s11434-016-1080-4}
  {\bibfield  {journal} {\bibinfo  {journal} {Sci. Bull.}\ }\textbf {\bibinfo
  {volume} {61}},\ \bibinfo {pages} {921} (\bibinfo {year} {2016})}\BibitemShut
  {NoStop}%
\bibitem [{\citenamefont {Petrovic}\ \emph
  {et~al.}(2001{\natexlab{b}})\citenamefont {Petrovic}, \citenamefont
  {Movshovich}, \citenamefont {Jaime}, \citenamefont {Pagliuso}, \citenamefont
  {Hundley}, \citenamefont {Sarrao}, \citenamefont {Fisk},\ and\ \citenamefont
  {Thompson}}]{Petrovic-CeIrIn5SC}%
  \BibitemOpen
  \bibfield  {author} {\bibinfo {author} {\bibfnamefont {C.}~\bibnamefont
  {Petrovic}}, \bibinfo {author} {\bibfnamefont {R.}~\bibnamefont
  {Movshovich}}, \bibinfo {author} {\bibfnamefont {M.}~\bibnamefont {Jaime}},
  \bibinfo {author} {\bibfnamefont {P.~G.}\ \bibnamefont {Pagliuso}}, \bibinfo
  {author} {\bibfnamefont {M.~F.}\ \bibnamefont {Hundley}}, \bibinfo {author}
  {\bibfnamefont {J.~L.}\ \bibnamefont {Sarrao}}, \bibinfo {author}
  {\bibfnamefont {Z.}~\bibnamefont {Fisk}}, \ and\ \bibinfo {author}
  {\bibfnamefont {J.~D.}\ \bibnamefont {Thompson}},\ }\href {\doibase
  10.1209/epl/i2001-00161-8} {\bibfield  {journal} {\bibinfo  {journal} {EPL}\
  }\textbf {\bibinfo {volume} {53}},\ \bibinfo {pages} {354} (\bibinfo {year}
  {2001}{\natexlab{b}})}\BibitemShut {NoStop}%
\bibitem [{\citenamefont {Takaesu}\ \emph {et~al.}(2011)\citenamefont
  {Takaesu}, \citenamefont {Aso}, \citenamefont {Tamaki}, \citenamefont {Hedo},
  \citenamefont {Nakama}, \citenamefont {Uchima}, \citenamefont {Ishikawa},
  \citenamefont {Deguchi},\ and\ \citenamefont {Sato}}]{Takaesu-CeIrIn5Pre}%
  \BibitemOpen
  \bibfield  {author} {\bibinfo {author} {\bibfnamefont {Y.}~\bibnamefont
  {Takaesu}}, \bibinfo {author} {\bibfnamefont {N.}~\bibnamefont {Aso}},
  \bibinfo {author} {\bibfnamefont {Y.}~\bibnamefont {Tamaki}}, \bibinfo
  {author} {\bibfnamefont {M.}~\bibnamefont {Hedo}}, \bibinfo {author}
  {\bibfnamefont {T.}~\bibnamefont {Nakama}}, \bibinfo {author} {\bibfnamefont
  {K.}~\bibnamefont {Uchima}}, \bibinfo {author} {\bibfnamefont
  {Y.}~\bibnamefont {Ishikawa}}, \bibinfo {author} {\bibfnamefont
  {K.}~\bibnamefont {Deguchi}}, \ and\ \bibinfo {author} {\bibfnamefont
  {N.~K.}\ \bibnamefont {Sato}},\ }\href {\doibase
  10.1088/1742-6596/273/1/012058} {\bibfield  {journal} {\bibinfo  {journal}
  {J. Phys: Conference Series}\ }\textbf {\bibinfo {volume} {273}},\ \bibinfo
  {pages} {012058} (\bibinfo {year} {2011})}\BibitemShut {NoStop}%
\bibitem [{\citenamefont {L\"{o}hneysen}\ \emph {et~al.}(2001)\citenamefont
  {L\"{o}hneysen}, \citenamefont {Pfleiderer}, \citenamefont {Pietrus},
  \citenamefont {Stockert},\ and\ \citenamefont
  {Will}}]{Lohneysen-CeCu6Au2001}%
  \BibitemOpen
  \bibfield  {author} {\bibinfo {author} {\bibfnamefont {H.~v.}\ \bibnamefont
  {L\"{o}hneysen}}, \bibinfo {author} {\bibfnamefont {C.}~\bibnamefont
  {Pfleiderer}}, \bibinfo {author} {\bibfnamefont {T.}~\bibnamefont {Pietrus}},
  \bibinfo {author} {\bibfnamefont {O.}~\bibnamefont {Stockert}}, \ and\
  \bibinfo {author} {\bibfnamefont {B.}~\bibnamefont {Will}},\ }\href {\doibase
  10.1103/PhysRevB.63.134411} {\bibfield  {journal} {\bibinfo  {journal} {Phys.
  Rev. B}\ }\textbf {\bibinfo {volume} {63}},\ \bibinfo {pages} {134411}
  (\bibinfo {year} {2001})}\BibitemShut {NoStop}%
\bibitem [{\citenamefont {Park}\ \emph {et~al.}(2008)\citenamefont {Park},
  \citenamefont {Sidorov}, \citenamefont {Ronning}, \citenamefont {Zhu},
  \citenamefont {Tokiwa}, \citenamefont {Lee}, \citenamefont {Bauer},
  \citenamefont {Movshovich}, \citenamefont {Sarrao},\ and\ \citenamefont
  {Thompson}}]{Park-CeRhIn5Iso}%
  \BibitemOpen
  \bibfield  {author} {\bibinfo {author} {\bibfnamefont {T.}~\bibnamefont
  {Park}}, \bibinfo {author} {\bibfnamefont {V.~A.}\ \bibnamefont {Sidorov}},
  \bibinfo {author} {\bibfnamefont {F.}~\bibnamefont {Ronning}}, \bibinfo
  {author} {\bibfnamefont {J.~X.}\ \bibnamefont {Zhu}}, \bibinfo {author}
  {\bibfnamefont {Y.}~\bibnamefont {Tokiwa}}, \bibinfo {author} {\bibfnamefont
  {H.}~\bibnamefont {Lee}}, \bibinfo {author} {\bibfnamefont {E.~D.}\
  \bibnamefont {Bauer}}, \bibinfo {author} {\bibfnamefont {R.}~\bibnamefont
  {Movshovich}}, \bibinfo {author} {\bibfnamefont {J.~L.}\ \bibnamefont
  {Sarrao}}, \ and\ \bibinfo {author} {\bibfnamefont {J.~D.}\ \bibnamefont
  {Thompson}},\ }\href {\doibase 10.1038/nature07431} {\bibfield  {journal}
  {\bibinfo  {journal} {Nature}\ }\textbf {\bibinfo {volume} {456}},\ \bibinfo
  {pages} {366} (\bibinfo {year} {2008})}\BibitemShut {NoStop}%
\bibitem [{\citenamefont {Custers}\ \emph {et~al.}(2012)\citenamefont
  {Custers}, \citenamefont {Lorenzer}, \citenamefont {M\"{u}ller},
  \citenamefont {Prokofiev}, \citenamefont {Sidorenko}, \citenamefont
  {Winkler}, \citenamefont {Strydom}, \citenamefont {Shimura}, \citenamefont
  {Sakakibara}, \citenamefont {Yu}, \citenamefont {Si},\ and\ \citenamefont
  {Paschen}}]{Custers-Ce3Pd20Si6QCP}%
  \BibitemOpen
  \bibfield  {author} {\bibinfo {author} {\bibfnamefont {J.}~\bibnamefont
  {Custers}}, \bibinfo {author} {\bibfnamefont {K.-A.}\ \bibnamefont
  {Lorenzer}}, \bibinfo {author} {\bibfnamefont {M.}~\bibnamefont
  {M\"{u}ller}}, \bibinfo {author} {\bibfnamefont {A.}~\bibnamefont
  {Prokofiev}}, \bibinfo {author} {\bibfnamefont {A.}~\bibnamefont
  {Sidorenko}}, \bibinfo {author} {\bibfnamefont {H.}~\bibnamefont {Winkler}},
  \bibinfo {author} {\bibfnamefont {A.~M.}\ \bibnamefont {Strydom}}, \bibinfo
  {author} {\bibfnamefont {Y.}~\bibnamefont {Shimura}}, \bibinfo {author}
  {\bibfnamefont {T.}~\bibnamefont {Sakakibara}}, \bibinfo {author}
  {\bibfnamefont {R.}~\bibnamefont {Yu}}, \bibinfo {author} {\bibfnamefont
  {Q.}~\bibnamefont {Si}}, \ and\ \bibinfo {author} {\bibfnamefont
  {S.}~\bibnamefont {Paschen}},\ }\href {\doibase 10.1038/nmat3214} {\bibfield
  {journal} {\bibinfo  {journal} {Nat. Mater.}\ }\textbf {\bibinfo {volume}
  {11}},\ \bibinfo {pages} {189} (\bibinfo {year} {2012})}\BibitemShut
  {NoStop}%
\bibitem [{\citenamefont {Luo}\ \emph {et~al.}(2014)\citenamefont {Luo},
  \citenamefont {Pourovskii}, \citenamefont {Rowley}, \citenamefont {Li},
  \citenamefont {Feng}, \citenamefont {Georges}, \citenamefont {Dai},
  \citenamefont {Cao}, \citenamefont {Xu}, \citenamefont {Si},\ and\
  \citenamefont {Ong}}]{LuoY-CeNiAsOQCP}%
  \BibitemOpen
  \bibfield  {author} {\bibinfo {author} {\bibfnamefont {Y.}~\bibnamefont
  {Luo}}, \bibinfo {author} {\bibfnamefont {L.}~\bibnamefont {Pourovskii}},
  \bibinfo {author} {\bibfnamefont {S.~E.}\ \bibnamefont {Rowley}}, \bibinfo
  {author} {\bibfnamefont {Y.}~\bibnamefont {Li}}, \bibinfo {author}
  {\bibfnamefont {C.}~\bibnamefont {Feng}}, \bibinfo {author} {\bibfnamefont
  {A.}~\bibnamefont {Georges}}, \bibinfo {author} {\bibfnamefont
  {J.}~\bibnamefont {Dai}}, \bibinfo {author} {\bibfnamefont {G.}~\bibnamefont
  {Cao}}, \bibinfo {author} {\bibfnamefont {Z.}~\bibnamefont {Xu}}, \bibinfo
  {author} {\bibfnamefont {Q.}~\bibnamefont {Si}}, \ and\ \bibinfo {author}
  {\bibfnamefont {N.~P.}\ \bibnamefont {Ong}},\ }\href {\doibase
  10.1038/nmat3991} {\bibfield  {journal} {\bibinfo  {journal} {Nat. Mater.}\
  }\textbf {\bibinfo {volume} {13}},\ \bibinfo {pages} {777} (\bibinfo {year}
  {2014})}\BibitemShut {NoStop}%
\bibitem [{\citenamefont {Suzuki}\ \emph {et~al.}(1990)\citenamefont {Suzuki},
  \citenamefont {Kwon}, \citenamefont {Ozeki}, \citenamefont {Haga},\ and\
  \citenamefont {Kasuya}}]{Suzuki-CeAsAFM}%
  \BibitemOpen
  \bibfield  {author} {\bibinfo {author} {\bibfnamefont {T.}~\bibnamefont
  {Suzuki}}, \bibinfo {author} {\bibfnamefont {Y.~S.}\ \bibnamefont {Kwon}},
  \bibinfo {author} {\bibfnamefont {S.}~\bibnamefont {Ozeki}}, \bibinfo
  {author} {\bibfnamefont {Y.}~\bibnamefont {Haga}}, \ and\ \bibinfo {author}
  {\bibfnamefont {T.}~\bibnamefont {Kasuya}},\ }\href {\doibase
  10.1016/S0304-8853(10)80178-7} {\bibfield  {journal} {\bibinfo  {journal} {J.
  Mag. Mag. Mater.}\ }\textbf {\bibinfo {volume} {90-91}},\ \bibinfo {pages}
  {493} (\bibinfo {year} {1990})}\BibitemShut {NoStop}%
\bibitem [{\citenamefont {Gegenwart}\ \emph {et~al.}(2008)\citenamefont
  {Gegenwart}, \citenamefont {Si},\ and\ \citenamefont
  {Steglich}}]{Gegenwart2008}%
  \BibitemOpen
  \bibfield  {author} {\bibinfo {author} {\bibfnamefont {P.}~\bibnamefont
  {Gegenwart}}, \bibinfo {author} {\bibfnamefont {Q.}~\bibnamefont {Si}}, \
  and\ \bibinfo {author} {\bibfnamefont {F.}~\bibnamefont {Steglich}},\ }\href
  {\doibase 10.1038/nphys892} {\bibfield  {journal} {\bibinfo  {journal} {Nat.
  Phys.}\ }\textbf {\bibinfo {volume} {4}},\ \bibinfo {pages} {186} (\bibinfo
  {year} {2008})}\BibitemShut {NoStop}%
\bibitem [{\citenamefont {Br\"uning}\ \emph {et~al.}(2008)\citenamefont
  {Br\"uning}, \citenamefont {Krellner}, \citenamefont {Baenitz}, \citenamefont
  {Jesche}, \citenamefont {Steglich},\ and\ \citenamefont
  {Geibel}}]{Bruning-CeFePOFMHF}%
  \BibitemOpen
  \bibfield  {author} {\bibinfo {author} {\bibfnamefont {E.~M.}\ \bibnamefont
  {Br\"uning}}, \bibinfo {author} {\bibfnamefont {C.}~\bibnamefont {Krellner}},
  \bibinfo {author} {\bibfnamefont {M.}~\bibnamefont {Baenitz}}, \bibinfo
  {author} {\bibfnamefont {A.}~\bibnamefont {Jesche}}, \bibinfo {author}
  {\bibfnamefont {F.}~\bibnamefont {Steglich}}, \ and\ \bibinfo {author}
  {\bibfnamefont {C.}~\bibnamefont {Geibel}},\ }\href {\doibase
  10.1103/PhysRevLett.101.117206} {\bibfield  {journal} {\bibinfo  {journal}
  {Phys. Rev. Lett.}\ }\textbf {\bibinfo {volume} {101}},\ \bibinfo {pages}
  {117206} (\bibinfo {year} {2008})}\BibitemShut {NoStop}%
\bibitem [{\citenamefont {Luo}\ \emph {et~al.}(2010)\citenamefont {Luo},
  \citenamefont {Li}, \citenamefont {Jiang}, \citenamefont {Dai}, \citenamefont
  {Cao},\ and\ \citenamefont {Xu}}]{LuoY-CeFeAsPO}%
  \BibitemOpen
  \bibfield  {author} {\bibinfo {author} {\bibfnamefont {Y.}~\bibnamefont
  {Luo}}, \bibinfo {author} {\bibfnamefont {Y.}~\bibnamefont {Li}}, \bibinfo
  {author} {\bibfnamefont {S.}~\bibnamefont {Jiang}}, \bibinfo {author}
  {\bibfnamefont {J.}~\bibnamefont {Dai}}, \bibinfo {author} {\bibfnamefont
  {G.}~\bibnamefont {Cao}}, \ and\ \bibinfo {author} {\bibfnamefont {Z.-a.}\
  \bibnamefont {Xu}},\ }\href {\doibase 10.1103/PhysRevB.81.134422} {\bibfield
  {journal} {\bibinfo  {journal} {Phys. Rev. B}\ }\textbf {\bibinfo {volume}
  {81}},\ \bibinfo {pages} {134422} (\bibinfo {year} {2010})}\BibitemShut
  {NoStop}%
\bibitem [{\citenamefont {Wang}\ \emph {et~al.}(2017)\citenamefont {Wang},
  \citenamefont {Fu}, \citenamefont {Sun}, \citenamefont {Liu}, \citenamefont
  {Yi}, \citenamefont {Yi}, \citenamefont {Luo}, \citenamefont {Dai},
  \citenamefont {Liu}, \citenamefont {Matsushita}, \citenamefont {Yamaura},
  \citenamefont {Lu}, \citenamefont {Cheng}, \citenamefont {feng Yang},
  \citenamefont {Shi},\ and\ \citenamefont {Luo}}]{WangL-CeCo2Ga8}%
  \BibitemOpen
  \bibfield  {author} {\bibinfo {author} {\bibfnamefont {L.}~\bibnamefont
  {Wang}}, \bibinfo {author} {\bibfnamefont {Z.}~\bibnamefont {Fu}}, \bibinfo
  {author} {\bibfnamefont {J.}~\bibnamefont {Sun}}, \bibinfo {author}
  {\bibfnamefont {M.}~\bibnamefont {Liu}}, \bibinfo {author} {\bibfnamefont
  {W.}~\bibnamefont {Yi}}, \bibinfo {author} {\bibfnamefont {C.}~\bibnamefont
  {Yi}}, \bibinfo {author} {\bibfnamefont {Y.}~\bibnamefont {Luo}}, \bibinfo
  {author} {\bibfnamefont {Y.}~\bibnamefont {Dai}}, \bibinfo {author}
  {\bibfnamefont {G.}~\bibnamefont {Liu}}, \bibinfo {author} {\bibfnamefont
  {Y.}~\bibnamefont {Matsushita}}, \bibinfo {author} {\bibfnamefont
  {K.}~\bibnamefont {Yamaura}}, \bibinfo {author} {\bibfnamefont
  {L.}~\bibnamefont {Lu}}, \bibinfo {author} {\bibfnamefont {J.-G.}\
  \bibnamefont {Cheng}}, \bibinfo {author} {\bibfnamefont {Y.}~\bibnamefont
  {feng Yang}}, \bibinfo {author} {\bibfnamefont {Y.}~\bibnamefont {Shi}}, \
  and\ \bibinfo {author} {\bibfnamefont {J.}~\bibnamefont {Luo}},\ }\href
  {\doibase 10.1038/s41535-017-0040-9} {\bibfield  {journal} {\bibinfo
  {journal} {npj Quantum Materials}\ }\textbf {\bibinfo {volume} {2}},\
  \bibinfo {pages} {36} (\bibinfo {year} {2017})}\BibitemShut {NoStop}%
\bibitem [{\citenamefont {Brown}(1981)}]{Brown-BVP1981}%
  \BibitemOpen
  \bibfield  {author} {\bibinfo {author} {\bibfnamefont {I.~D.}\ \bibnamefont
  {Brown}},\ }\href@noop {} {\emph {\bibinfo {title} {{Sturcture and Bonding in
  Crystals}}}},\ Vol.~\bibinfo {volume} {2}\ (\bibinfo  {publisher} {Academic
  Press, New York},\ \bibinfo {year} {1981})\BibitemShut {NoStop}%
\bibitem [{\citenamefont {Brown}\ and\ \citenamefont
  {Altermatt}(1985)}]{Brown-BVP1985}%
  \BibitemOpen
  \bibfield  {author} {\bibinfo {author} {\bibfnamefont {I.~D.}\ \bibnamefont
  {Brown}}\ and\ \bibinfo {author} {\bibfnamefont {D.}~\bibnamefont
  {Altermatt}},\ }\href {\doibase 10.1107/S0108768185002063} {\bibfield
  {journal} {\bibinfo  {journal} {Acta Cryst. B}\ }\textbf {\bibinfo {volume}
  {41}},\ \bibinfo {pages} {244} (\bibinfo {year} {1985})}\BibitemShut
  {NoStop}%
\bibitem [{\citenamefont {Brese}\ and\ \citenamefont
  {O'Keeffe}(1991)}]{Brese-BVP1991}%
  \BibitemOpen
  \bibfield  {author} {\bibinfo {author} {\bibfnamefont {N.~E.}\ \bibnamefont
  {Brese}}\ and\ \bibinfo {author} {\bibfnamefont {M.}~\bibnamefont
  {O'Keeffe}},\ }\href {\doibase 10.1107/S0108768190011041} {\bibfield
  {journal} {\bibinfo  {journal} {Acta Cryst. B}\ }\textbf {\bibinfo {volume}
  {47}},\ \bibinfo {pages} {192} (\bibinfo {year} {1991})}\BibitemShut
  {NoStop}%
\bibitem [{\citenamefont {Boutron}(1973)}]{Boutron-PMchi}%
  \BibitemOpen
  \bibfield  {author} {\bibinfo {author} {\bibfnamefont {P.}~\bibnamefont
  {Boutron}},\ }\href {\doibase 10.1103/PhysRevB.7.3226} {\bibfield  {journal}
  {\bibinfo  {journal} {Phys. Rev. B}\ }\textbf {\bibinfo {volume} {7}},\
  \bibinfo {pages} {3226} (\bibinfo {year} {1973})}\BibitemShut {NoStop}%
\end{thebibliography}

\end{document}